# MICROFABRICATION TECHNOLOGY FOR LARGE LEKID ARRAYS: FROM NIKA2 TO FUTURE APPLICATIONS

J. Goupy 1, A. Adane 2, A. Benoit 1, O. Bourrion 3, M. Calvo 1, A. Catalano 3-1, G. Coiffard 2, C. Hoarau 1, S. Leclercq 2, H. Le Sueur 4, J. Macias-Perez 3, A. Monfardini 1-3, I. Peck 5, K. Schuster 3

*1 Institut Néel & Université Joseph Fourier, CNRS, Grenoble, France*
*2 Institut de Radio Astronomie Millimétrique (IRAM), Grenoble, France*
*3 LPSC, Université Grenoble-Alpes, CNRS/IN2P3, Grenoble, France*
*4 Centre de Spectrométrie Nucléaire et de Spectrométrie de Masse, Paris, France*
*5 PTA/CIME/NANOTECH/INPG, Grenoble, France*

**Abstract:** The Lumped Element Kinetic Inductance Detectors (LEKID) demonstrated full maturity in the NIKA (New IRAM KID Arrays) instrument. These results allow directly comparing LEKID performance with other competing technologies (TES, doped silicon) in the mm and sub-mm range. A continuing effort is ongoing to improve the microfabrication technologies and concepts in order to satisfy the requirements of new instruments. More precisely, future satellites dedicated to CMB (Cosmic Microwave Background) studies will require the same focal plane technology to cover, at least, the frequency range of 60 to 600 GHz. Aluminium LEKID developed for NIKA have so far demonstrated, under real telescope conditions, performance approaching photon-noise limitation in the band 120-300 GHz. By implementing superconducting bi-layers we recently demonstrated LEKID arrays working in the range 80-120 GHz and with sensitivities approaching the goals for CMB missions. NIKA itself (350 pixels) is followed by a more ambitious project requiring several thousands (3000-5000) pixels. NIKA2 has been installed in October 2015 at the IRAM 30-m telescope. We will describe in detail the technological improvements that allowed a relatively harmless 10-fold up-scaling in pixels count without degrading the initial sensitivity. In particular we will briefly describe a solution to simplify the difficult fabrication step linked to the slot-line propagation mode in coplanar waveguide.

1. Introduction

The KID technology is ready to provide detectors with sensitivity near photon noise limit under real observing conditions. Many instruments in millimetric astronomy, ground or space-based, move forwards the KID technology. The relative simplicity of the KID technology is a strength compared to concurrent technologies. The New IRAM KID Arrays (NIKA) instrument has been open to the 30 meters IRAM telescope observers in the period 2014-2015. The observing campaigns have demonstrated state-of-the-art sensitivities at both 1.25 and 2 mm wavelengths and good photometry performances, giving a bulk of astrophysical results, e.g. [1]. The focal planes are made of 356 pixels LEKID split into two arrays. The pixels are multiplexed on one feedline per array. A thin layer (~18 nm) of aluminium is used for these arrays. The detectors and part of the optics are cooled down to 100 mK in a dilution cryostat. In the 120-300 GHz band, the sensitivity is close to the photon noise limit [2]. In October 2015 the NIKA instrument has been replaced by the next generation NIKA2 camera for continuum and polarization measurements using a total of 3500 pixels shared between three focal planes operating at 1.15 mm and 2 mm [3].

For a complete review of the KIDs theory, we suggest [4] [5] [6].

In this paper we review different upgrades that we have developed for future detectors. First, we define the classical technological process fabrication for the NIKA2 detectors. In the second part, we discuss improvements of KID for NIKA2 detectors. In the third part, we present the new technological solution developed based on titanium / aluminium bilayer to address the so called "3 mm band" (80-120 GHz).

2. The standard technological process for NIKA2

The NIKA2 standard fabrication process is derived from the NIKA one. In particular, the geometry of the pixels is almost identical. The NIKA2 arrays contain ~1000 pixels for the 2 mm band and ~2000 pixels for the 1.15 mm band, so a factor about 10 times more compared to NIKA. These thousand pixels arrays are fabricated on single 4 inch silicon wafers. Only 4 or 8 feedlines are used to read all the pixels in respectively the 2 and 1.15 mm bands. The silicon monocrystalline [100] wafer is a high purity bulk (>1000 Ohm.cm), chosen with a thickness corresponding to $\lambda/4$ of the wavelength to be detected (2 mm and 1.15 mm).

The technological process is pretty simple. The most critical point in LEKID fabrication is related to the defects on the array and in particular in the feedlines. In the fabrication process, particular care is adopted concerning the cleanness of the wafer, the cleaning between process steps and the quality of the thin aluminium layer. For a characteristic impedance of 50 Ohm, the width of the commonly used coplanar waveguide (CPW) is 40 µm and the gap 18 µm. The typical length of one feedline is nearly 0.5 meter. The yield in fabrication of this kind of large arrays in our cleanroom is around 50%. Switching to alternative feeding strategies described in the next paragraph, the yield increases up to 80%.

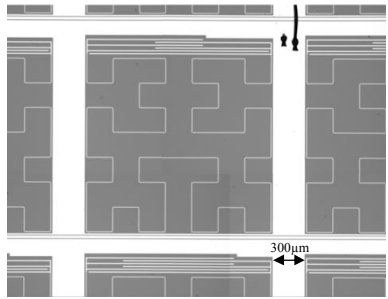

**Fig. 1** Picture of 2 mm pixel with CPW feedline (gap 18 µm / line 40 µm)

The first step is the cleaning of the 4 inch silicon wafer. After the Acetone and Isopropanol baths, the wafer is inserted in the evaporator machine. Argon plasma is adopted for etching the native $SiO_2$ layer. A thin layer of aluminium is then evaporated without breaking the vacuum. The thickness is to be adjusted according to the design and to match the typical incoming power per pixel. We have determined an optimum at 18 nm in the case of NIKA2. The evaporation is achieved at low rate deposition (0.1 nm/s), at 1.2E-7 Torr and keeping the substrate at room temperature.

Before the patterning of the resonators, the wafer is baked at 180 °C during 2 minutes in order to dehydrate the surface. Afterward, a positive photoresist is spinned and then baked at 110 °C during 2 minutes. The resist layer is exposed to 365 nm radiation through a mask. The resist is then developed and post-baked. The surface of the aluminium not protected by the resist is finally wet etched by a specific solution based on diluted $H_3PO_4$ ( Aluminum Etch 1960 W/AES). The obtained circuit is packaged after a careful visual checking. The array is diced to be integrated in a dedicated holder equipped with SMA connectors.

In the case of CPW feed arrays, incoming photons pass through the substrate before reaching the KID. The thickness of the wafer is optimized according to the radiation to be detected (example: 300 µm for the 2 mm band). A n-index matching layer on the back of the silicon wafer minimizes the reflections at the Si-vacuum boundary (see Fig. 2) [7]. With a micro-mechanical saw, we have drawn a grid of grooves, with a width of 100 μm and a depth of 225 μm.

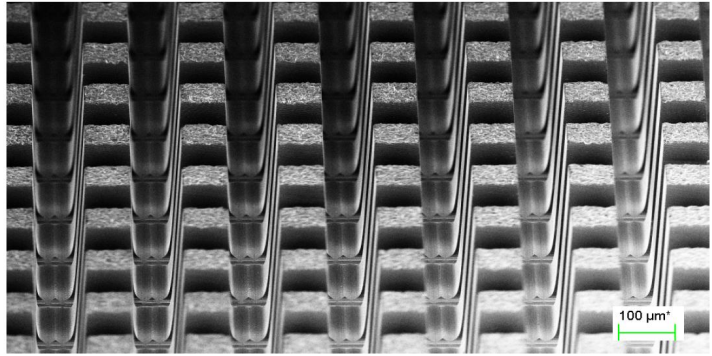

**Fig. 2** The back-face of the wafer after the optical adaptation treatment by dicing saw. The effective index achieved by removing 70% of the silicon is n ~ 2.

**3. Microstrip arrays for NIKA2**

CPW-fed resonator arrays suffer traditionally from slot-line modes that lead to large coupling dispersion across the matrix. Losses in CPW feedlines in our arrays can be neglected [10]. To attenuate this effect, we usually solder micro-wire-bondings between the two sides of the ground plane. This step is risky and time-consuming. In NIKA2, micro-bridges were realized by aluminium bondings ~ 17 μm in diameter with a dedicated machine (see Fig. 3). We have thus investigated alternative single-mode feeds. The first tests with coplanar strip-line (CPS) and new pixel geometry (described later in this section) shows that the placement of resonances in 100 pixels arrays is less shuffled (see Fig. 5). In this test, the pixel geometry is almost the same as the one shown in Fig.4, and the microstrip is replaced by a CPS feedline. Unfortunately, with the CPS, the homogeneity of the coupling quality factors of resonances was to dispersive (> 100%).

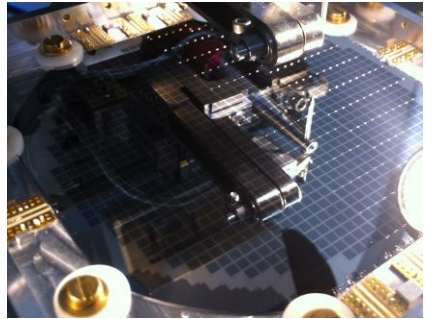

**Fig. 3** CPW 2 mm NIKA2 array undergoing the wire-bonding step. ~ 400 aluminium micro-wires of 17 μm of diameter are soldered in bridge shape above CPW feedline each 3 pixels with a dedicated machine.

The next solution was adopting a microstrip design (see Fig. 4). For the microstrip and CPS designs, the ground plane is removed compared to the CPW one. To reduce the electromagnetic cross-coupling between pixels and further decouple the resonator from the feedline, a shield (a "loop") is added around each resonator. This trick allowed to solve at the same time two key electromagnetic issues. Microstrip design (see Fig. 4) is today integrated on NIKA2 arrays. Both the 1.15 mm arrays mounted in NIKA2 are in fact realized in microstrip technology. We have obtained the same pixels efficiency with the microstrip design as with the original CPW design [3]. The fabrication process is the same compared to CPW classical process, except that an aluminium ground plan is added on the bottom side of the wafer at the end of the process.

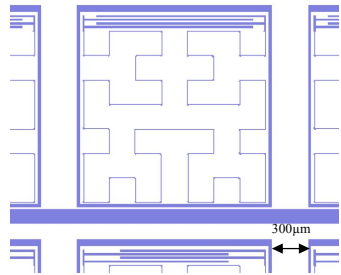

**Fig. 4** Geometry for a 2 mm microstrip pixel with EM shield ("loop" geometry)

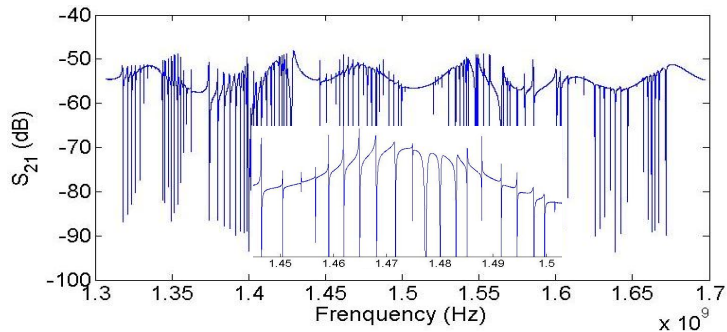

**Fig. 5** Frequency sweep of one feedline with 100 NIKA2 pixels ("loop" geometry). It consists on five blocks of 20 pixels at a distance of 16 MHz each.

Slight variations of thickness of the aluminium layer and lithography-etching small imperfections are known to induce important frequency shuffling due to the kinetic and geometric inductance variations. This subject is discussed in a recent study [8]. We stress that the entire NIKA2 production has been achieved using general purpose machines open to the other users. The use of multi-purpose machines open to many different users can have a negative effect on the quality of the produced films that can be contaminated by other materials evaporated in the same machine. Furthermore, the intensive usage of such equipments can lead to malfunctioning and therefore lower productivities. Specific deposition techniques (e.g. co-focal sputtering) and dedicated machines will probably be mandatory in the case of space-qualified detectors production, to increase the uniformity and purity of the layer evaporated and to decrease the number of defective pixels in large arrays.

**4. The 80-120 GHz band with Ti/Al bilayer**

The future CMB missions from space will address a multi-band instrument with a large spectral coverage (~60-600 GHz). With aluminium KID, we have recently demonstrated that it is possible to meet the requirements for the band 120-350 GHz. For the lowest band (60-120 GHz), pure and thin (<20 nm) aluminium cannot be used due to its superconducting gap ($2\Delta Al \sim$ 110 GHz). For the upper band (350 to 600 GHz), we are investigating different solutions, including again pure aluminium. Here, we highlight the titanium / aluminium technology. With this bilayer we have recently demonstrated very good performances for the band 80-120 GHz [9].

This technology is based on the proximity effect between the aluminium ($T_c$ ~ 1.4 K) and the titanium ($T_c$ ~ 0.45 K). We have fabricated 25 pixels and 132 pixels arrays with 10 nm titanium / 25 nm aluminium bilayer adopting the NIKA pixels design. In terms of technological process, we follow the classical NIKA process until the wet etch step. In the bilayer case, after the aluminium etch the wafer is dipped in HF bath (diluted at 0.2%, during 15 seconds) to remove the titanium left exposed.

The bilayer, as expected, exhibits a critical temperature around 0.9 K. Even before elaborating an optimised design for this band, we have achieved good results on the spectral absorption band (see Fig. 6a) and measured a sensitivity within only a factor ~2 compared to the NEP (Noise Equivalent Power) goal for future CMB missions (see Fig. 6b).

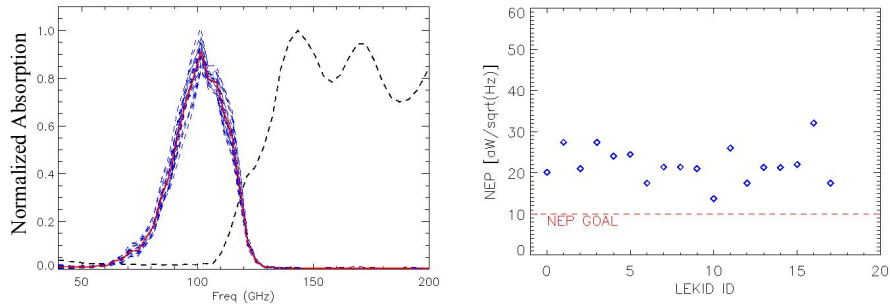

**Fig. 6**: *Left*: Normalized spectral absorption of individual detectors (blue dashed lines) and averaged (red line) of the Ti-Al array. We can see the impact of the high-frequency cut-off at 120 GHz. The black dash line is the classical spectral absorption for 2 mm aluminium arrays (without high-frequency cut-off). *Right*: individual pixel sensitivities (blue diamonds) measured with an optical load of 0.3 pW compared to the reference goal (dashed red line). The reference goal corresponds to twice the in-space photon noise level per pixel at 100 GHz with a bandwidth $\Delta\nu/\nu = 0.3$.

**5. Conclusions**

We have shown the relative simplicity of the fabrication of KID large (thousands pixels) arrays. Today the aluminium KID technology has demonstrated state-of-the-art performance at millimeter wavelengths [2] [3]. We are continuously improving the KID design and technology to meet future requests of detectors for space and balloon CMB missions. We are working to further reduce the frequency shuffling of the resonances in large arrays. We have addressed spectral bands of absorption not accessible by aluminium technology with the development of the bilayer titanium / aluminium KID technology.

**Acknowledgements:** This work has been performed at the "Plateforme Technologique Amont" (PTA) of Grenoble, with the financial support of the Labex "Focus", the CNES and the ANR. We acknowledge the support of the cryogenics and electronics groups at the Institute NEEL.